\documentstyle[12pt,epsf,a4wide]{article}

\begin{document}
\renewcommand{\thefootnote}{\fnsymbol{footnote}}
\makebox[2cm]{}\\[-1in]
\begin{flushright}
\begin{tabular}{l}
TUM--T31--73/94\\
hep-ph/9408385n
\end{tabular}
\end{flushright}
\vskip0.4cm
\begin{center}
{\Large\bf
Radiative Corrections to Nonleptonic Inclusive\\[4pt]
B Decays and the Semileptonic Branching Ratio\\[4pt]
of B Mesons\footnotemark}
\footnotetext{Invited talk given at QCD 94, Montpellier, France, 7--13
July 1994; to appear in the Proceedings}

\bigskip

Patricia Ball

\bigskip

{\em Physik--Department, TU M\"{u}nchen, D--85747 Garching, Germany}

\bigskip

{\em August 29, 1994}

\bigskip

{\bf Abstract:\\[5pt]}
\parbox[t]{\textwidth}{\small
We calculate the radiative corrections to the nonleptonic inclusive B
decay mode $b\to c ud$ taking into account the charm quark mass. The
corrections increase the decay rate by (4--8)\%, depending on the
renormalization point. Using these results, we obtain an improved
theoretical prediction for the semileptonic branching ratio of B
mesons. This talk relies on work done in collaboration with E. Bagan,
V.M.\ Braun and P. Gosdzinsky.}
\end{center}

\section{Introduction}
Owing to the newly developed tool of an expansion in the inverse heavy
quark mass \cite{general}, the theoretical description of weak inclusive
decays
of heavy mesons now rests on a more solid ground than ever. Since in
such decays the energy release is large compared to the masses of
the final state particles, the process takes place essentially at small
distances and in leading order in the heavy quark expansion (HQE) is
described by the underlying quark decay process.
Hadronic corrections only enter at second order in the HQE
and are of natural size $\sim 1\,{\rm GeV}^2/m_b^2\sim
5\%$ for B decays. Thus the accuracy of
theoretical predictions of hadronic quantities like, say, the
semileptonic branching ratio is not so much limited by the
necessarily incomplete knowledge of (non--perturbative)
hadronic matrix elements, but rather controlled by our knowledge of
{\em perturbative} corrections to the free quark decay.

Full corrections to ${\cal O}(\alpha_s)$ are
known for the semileptonic decay $b\to c e\nu$ \cite{CM78},
and for $b\to c u d$ in the limit of
massless final state quarks \cite{ACMP81}. Although it is known
that the exchange of gluons between quarks of unequal masses can yield
big effects (cf.\ the extreme case of an infinitely heavy heavy quark
investigated in \cite{BG92}), in existing analyses of the semileptonic
branching ratio of B mesons \cite{AP91} ,
$m_c$ was put zero in the radiative corrections to the
nonleptonic width. In order to improve the existing predictions,
we thus felt motivated to
calculate the radiative corrections to $b\to c u d$ with full account
for the c quark mass~\cite{tum67}.

\section{Method of Calculation}

Without going into too much details, I present a short outline of
 the calculation  done in \cite{tum67}. Our starting point was to
express the decay rate as imaginary part of the relevant
forward--scattering amplitude. We used $\overline{{\rm MS}}$ subtraction and
re\-gu\-la\-ri\-zed occurring ultraviolet divergencies with di\-men\-sio\-nal
regularization with anticommuting $\gamma_5$, often referred to as
na\"{\i}ve dimensional regularization (NDR). NDR is applicable if one uses
Fierz--transformations to relate diagrams with closed fermion loops,
which are ambiguous in NDR, to such diagrams that are well--defined in
NDR. As shown in \cite{BW90}, Fierz--transformations are valid
diagram by diagram only with the proper choice of the so--called
evanescent operators. We have verified that in the limit $m_c\to 0$
our procedure yields the same results as obtained in other schemes
 \cite{ACMP81}.

For the calculation of the imaginary parts of the forward--scattering
amplitudes, we used a rather conservative technique, namely applied Cutkosky
rules and regularized intermediate infra--red singularities by small
quark and gluon masses which allows phase--space
integration to be done in four dimensions.

\section{Results}

Since the complete formulas for $\Gamma(b\to c u d)$ are rather
involved, I present results only in form of plots.
Fig.~1 shows the effect of the non--vanishing c quark
mass on the quark decay rate
\begin{equation}
\Gamma(b\to c u d) = 3\Gamma_0 \eta(\mu) J(m_c/m_b,\mu).
\end{equation}
Here $\Gamma_0$ is the semileptonic tree--level decay rate, the factor
3 accounts for the colour enhancement in nonleptonic decays, $\mu$ is
the renormalization scale,
$\eta(\mu)$ contains the leading order QCD corrections,
$\eta(4.8\,{\rm GeV}) = 1.10$, and $J(m_c/m_b,\mu)$ gives the
next--to--leading order
corrections. In Fig.~1 the quantity $J(m_c/m_b,\mu)/J(0,\mu)$ is plotted as
function of $m_c/m_b$ for three different values of the
renormalization scale $\mu$. The grey bar denotes a conservative range
of ``physical'' quark masses. Finite c quark mass effects thus
constitute a $(4-8)\%$ increase of the decay rate $\Gamma(b\to c u d)$
with respect to the massless case.

Turning now to hadronic corrections, the HQE of $\Gamma(B\to
Xe\nu)$ involves to order $1/m_b^2$ two hadronic matrix elements:
\begin{eqnarray}
2m_B\lambda_1 & = & \langle\,B\,|\,\bar{b}_v (iD)^2
b_v\,|\,B\,\rangle, \nonumber\\
6m_B\lambda_2 & = & \langle\,B\,|\,\bar{b}_v \frac{g}{2}\,
\sigma_{\mu\nu} F^{\mu\nu}b_v\,|\,B\,\rangle,
\end{eqnarray}
where $b_v$ is defined as $b_v = e^{im_b v x}b(x)$, $b(x)$ being the
$b$ quark field in full QCD, $v_\mu$ is the four--velocity of
the B meson, $m_B$ its mass and $F^{\mu\nu}$ the
gluonic field--strength tensor.

Whereas $\lambda_2$ is directly related to the observable
spectrum of beautiful mesons,
\begin{equation}
\lambda_2 \approx \frac{1}{4}\,(m_{B^*}^2 - m_B^2)= 0.12\,{\rm GeV}^2,
\end{equation}
the quantity $\lambda_1$ is difficult to measure, cf.\
\cite{grozin}. Physically, $-\lambda_1/(2m_b)$ is
just the average kinetic energy of the b quark inside the meson.
In the present analysis I conform to the value $\lambda_1 =  -(0.6\pm
0.1)\,{\rm GeV}^2$ obtained from QCD sum rules \cite{BB94}. For a
discussion of the present status of $\lambda_1$, I refer to \cite{NPro}.

\begin{figure}[t]
\centerline{
\epsfxsize=0.45\textwidth
\epsfbox{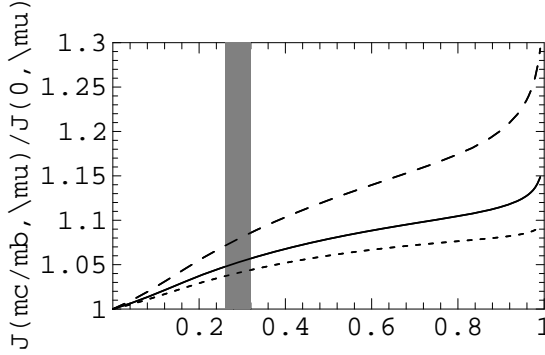}
}
\vspace{-0.4in}
\caption[]{The next--to-leading order corrections $J(m_c/m_b,\mu)$ to
$\Gamma(b\to c u d)$ as function of $m_c/m_b$, normalized to one at
$m_c=0$, for three different renormalization scales: solid line:
$\mu = m_b$, long dashes: $\mu =
m_b/2$, short dashes: $\mu = 2m_b$.}
\end{figure}
The semileptonic branching ratio is defined by
\begin{equation}\label{eq:BR}
B(B\to X e\nu) = \frac{\Gamma(B\to X e\nu)}{\Gamma_{tot}}
\end{equation}
with
\begin{equation}
\Gamma_{tot} = \sum_{\ell =
e,\,\mu,\,\tau}\Gamma(B\to X\ell\nu_{\ell}) + \Gamma(B\to
X_c)+ \Gamma(B\to X_{c\bar c}).
\end{equation}
The explicit formulas for the decay rates can be found in
\cite{general}. In evaluating (\ref{eq:BR}), it is crucial to minimize
the number of independent parameters. To this purpose, we
take advantage of the fact that the {\em difference} between
heavy quark masses is fixed in the framework of HQE:
\begin{equation}
m_b - m_c = m_B-m_D + \frac{\lambda_1+3\lambda_2}{2}\,\left(
\frac{1}{m_b} -\frac{1}{m_c} \right)
 + {\cal O} \left(\frac{1}{m^2}\right).\label{eq:gaehn}
\end{equation}
The only quantity remaining to be fixed is then $m_b$ or $m_c$. We
prefer to take $m_b$ from spectroscopy and choose the most
conservative range
\begin{equation}
4.5\,{\rm GeV}\leq m_b\leq 5.1\,{\rm GeV},
\end{equation}
which is broad enough to cover all uncertainties arising from the
renormalon ambiguity of the pole mass, cf.\ \cite{ren}.

In Fig.~2(a) we plot $B(B\to Xe\nu)$, Eq.\ (\ref{eq:BR}), as function of
$m_b$ with $\alpha_s(m_Z)=0.117$ \cite{bethke}. Here the widths are
expressed in terms of pole quark masses. Nevertheless the pole masses
are not the genuinely most suited ones for the analysis of weak
decays, cf.\ \cite{BN94}. Rewriting Eq.\ (\ref{eq:BR}) in terms of
running $\overline{\rm MS}$ masses, I obtain Fig.~2(b). Although
formally the difference to Fig.~2(a) is of higher order in $\alpha_s$,
it is clearly visible and illustrates the problem of {\em
scheme--dependence} that makes its appearance in any finite order
perturbative calculation.
\begin{figure}[t]
\centerline{
\epsfxsize=0.45\textwidth
\epsfbox{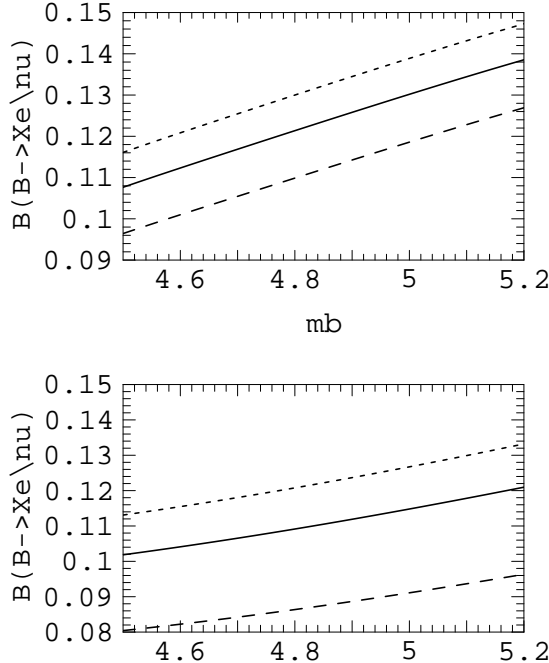}
}
\vspace{-0.4in}
\caption[]{(a) $B(B\to Xe\nu)$, Eq.\ (\protect{\ref{eq:BR}}), as
function of the pole mass $m_b$ with $m_b-m_c$
fixed by Eq.~(\protect{\ref{eq:gaehn}}). $m_b$ and $m_c$
are pole masses. The three lines have the same meaning as in Fig.~1.
(b) The same as (a), but with running masses $m(\mu)$. Yet, as in (a), the
abscissa is given in terms of the pole mass.}
\end{figure}

{}From Fig.~2 we obtain
\begin{equation}\label{eq:8}
B(B\to X e \nu) = (11.5\pm 1.3\pm 1.0)\%,
\end{equation}
where the first error combines uncertainties in $\alpha_s(m_Z)$,
$m_b$, the hadronic corrections, the
renormalization scale and unknown radiative corrections to $b\to ccs$.
The second error is a ``guestimate'' of the theoretical error due to
scheme--dependence. Eq.\ (\ref{eq:8})
has to be compared with previous theoretical analyses, which
consistently yielded $B(B\to Xe\nu)>12.5\%$
\cite{AP91}, the experimental world average $B(B\to Xe\nu)
= (10.43\pm 0.24)\%$ \cite{PartData} and the most recent CLEO measurement
$B(B^0\to X e \nu) = (10.9\pm 0.7\pm 1.1)\%$, \cite{Cleo}.
The combined effect of complete radiative
corrections, new results on $\alpha_s(m_Z)$ \cite{bethke} and the
consideration of different definitions of the quark mass
thus lowers the theoretical branching ratio, which now agrees with the
experimental one within the errors. A more detailed analysis is in
preparation \cite{tum68}.


\end{document}